\documentclass[twocolumn,twocolappendix]{aastex631}

\newcommand{\msol}{$M_\odot$}

\newcommand{\lya}{Ly$\alpha$}

\newcommand{\oiii}{[O{\small III}]}
\newcommand{\ha}{H$\alpha$}

\newcommand{\change}[1]{#1}

\usepackage{amsmath,amssymb,amsfonts}
\usepackage[caption=false]{subfig}
\usepackage[capitalise]{cleveref}

\begin{document}

\title{Quasar lifetime measurements from extended \lya\ nebulae at $z\sim 6$}

\author[0000-0001-8986-5235]{Dominika {\v D}urov{\v c}{\'i}kov{\'a}}
\affiliation{MIT Kavli Institute for Astrophysics and Space Research, 77 Massachusetts Avenue, Cambridge, 02139, Massachusetts, USA}
\affiliation{Department of Physics, Massachusetts Institute of Technology, 77 Massachusetts Avenue Cambridge, MA 02139}

\author[0000-0003-2895-6218]{Anna-Christina Eilers}
\affiliation{MIT Kavli Institute for Astrophysics and Space Research, 77 Massachusetts Avenue, Cambridge, 02139, Massachusetts, USA}
\affiliation{Department of Physics, Massachusetts Institute of Technology, 77 Massachusetts Avenue Cambridge, MA 02139}

\author[0000-0001-5492-4522]{Romain A.\ Meyer}
\affiliation{Department of Astronomy, University of Geneva, Chemin Pegasi 51, 1290 Versoix, Switzerland}

\author[0000-0002-6822-2254]{Emanuele Paolo Farina}
\affiliation{International Gemini Observatory/NSF NOIRLab, 670 N A’ohoku Place, Hilo, Hawai'i 96720, USA}

\author[0000-0002-2931-7824]{Eduardo Ba{\~n}ados}
\affiliation{Max Planck Institut f\"ur Astronomie, K\"onigstuhl 17, D-69117, Heidelberg, Germany}

\author[0000-0003-0821-3644]{Frederick B. Davies}
\affiliation{Max Planck Institut f\"ur Astronomie, K\"onigstuhl 17, D-69117, Heidelberg, Germany}

\author[0000-0002-7054-4332]{Joseph F.\ Hennawi}
\affiliation{Leiden Observatory, Leiden University, P.O. Box 9513, 2300 RA Leiden, The Netherlands}
\affiliation{Department of Physics, University of California, Santa Barbara, CA 93106, USA}

\author[0000-0002-5941-5214]{Chiara Mazzucchelli}
\affiliation{Instituto de Estudios Astrof\'{\i}sicos, Facultad de Ingenier\'{\i}a y Ciencias, Universidad Diego Portales, Avenida Ejercito Libertador 441, Santiago, Chile.}

\author[0000-0003-3769-9559]{Robert A.\ Simcoe}
\affiliation{MIT Kavli Institute for Astrophysics and Space Research, 77 Massachusetts Avenue, Cambridge, 02139, Massachusetts, USA}
\affiliation{Department of Physics, Massachusetts Institute of Technology, 77 Massachusetts Avenue Cambridge, MA 02139}

\author[0000-0003-4793-7880]{Fabian Walter}
\affiliation{Max Planck Institut f\"ur Astronomie, K\"onigstuhl 17, D-69117, Heidelberg, Germany}

\correspondingauthor{Dominika {\v D}urov{\v c}{\'i}kov{\'a}}
\email{dominika@mit.edu}



\begin{abstract}

The existence of billion-solar-mass black holes hosted in luminous quasars within the first gigayear of cosmic history poses a challenge to our understanding of supermassive black hole (SMBH) growth. The problem is further exacerbated by the very short quasar lifetimes of $t_{\rm Q}\lesssim 10^6$ years, as derived from the extent of their proximity zone (PZ) sizes observed in the quasars' rest-UV spectra. However, the quasar lifetime estimates based on the extents of the proximity zones may be underestimated, as time-variable obscuration effects might have limited the quasars' emission along our sightline in the past.
In this work, we present independent quasar lifetime measurements for six quasars at $z \sim 6$ leveraging the extended nebular emission \textit{perpendicular} to our line-of-sight. We use observations from the Very Large Telescope/Multi-Unit Spectroscopic Explorer (MUSE) to search for extended \lya\ emission in the circumgalactic medium around quasars with small proximity zones and estimate their lifetimes as the light travel time between the SMBH and the outer edge of the nebula. We find agreement between the independent lifetime estimates. For one object we find a proximate absorption system prematurely truncating the extent of the quasar's proximity zone, which thus results in an expected discrepancy between the lifetime estimates. Our results provide further evidence that the quasars' current accretion episode has only recently begun, challenging our models of SMBH growth. 

\end{abstract}



\section{Introduction} \label{sec:intro}

The growth of supermassive black holes (SMBHs) through accretion is accompanied by the release of copious amounts of radiation \citep{soltan_masses_1982, yu_observational_2002}, enabling their observation as quasars out to high redshifts \citep{fan_quasars_2023}. However, the discovery of billion-solar-mass black holes hosted in quasars at $z\gtrsim 6$ \citep[e.g.,][]{wu_ultraluminous_2015,banados_800-million-solar-mass_2018,wang_significantly_2020,wang_luminous_2021} presents a significant challenge to standard black hole growth models. Even under the assumption of continuous, Eddington-limited accretion, growing a SMBH from a $100$\,\msol\ seed to a billion solar masses would require nearly a billion years \citep{inayoshi_assembly_2020}. This timescale is comparable to the age of the Universe at these redshifts.

Maintaining such steady growth is challenging for many reasons, e.g. due to disruptions from BH feedback and supernova explosions \citep[e.g.][]{johnson_aftermath_2007,whalen_destruction_2008,zhou_modeling_2024}. A variety of alternative pathways have thus been proposed to allow a more rapid BH mass build-up over cosmic time, such as episodic accretion with phases of obscured growth and/or super-Eddington accretion \citep{davies_evidence_2019,satyavolu_need_2023}, growth driven by black hole mergers \citep[e.g.][]{volonteri_quasars_2006,tanaka_assembly_2009} or jet-assisted growth \citep{connor_uncovering_2024}.

The problem gets even more severe as high-redshift quasars appear to have only been active for a short amount of time. Short UV-luminous quasar lifetimes have been measured at $z\sim 6$ primarily from proximity zones (PZs), the regions of increased transmission blueward of \lya\ carved out into the intergalactic medium (IGM) by the quasars' ionizing radiation. Considering the balance of photoionization and recombination timescales for a hydrogen-rich IGM, \cite{eilers_implications_2017} used radiative transfer simulations to establish a relation between \lya\ PZ sizes and quasar lifetimes. Since then, such measurements of PZ sizes \citep{eilers_implications_2017,eilers_first_2018,eilers_detecting_2020,eilers_detecting_2021,morey_estimating_2021,satyavolu_new_2023}, as well as the modeling of damping wings \citep{davies_evidence_2019,durovcikova_chronicling_2024}, have revealed a population of quasars with surprisingly small proximity zones that imply lifetimes of $t_{\rm Q} \lesssim 10^6$ yr. These short SMBH growth timescales at $z\sim6$ have been supported by independent measurements of quasar duty cycles 
from clustering measurements \citep{pizzati_unified_2024,eilers_eiger_2024}.

Interestingly, one way to alleviate the tension between the long black hole growth timescales required by Eddington-limited accretion and the short timescales implied by quasar lifetime measurements at $z\sim6$ is if time-variable, sightline-dependent obscuration effects are present \citep{davies_evidence_2019,satyavolu_need_2023}. If luminous high-redshift quasars indeed underwent a significant portion of their SMBH growth in a partially obscuring medium, the UV emission along some sightlines could have been limited in the past, thus potentially imprinting smaller PZs or stronger damping wings in the observed spectra of these quasars \citep{davies_evidence_2019}. 
Likewise, the short quasar duty cycles measured via clustering studies of UV-luminous quasars at $z\sim6$ could be explained if the fraction of obscured quasars in the early universe is high \citep{eilers_eiger_2024}. 

To investigate the role that obscured growth plays in the assembly of these early SMBHs, we apply an alternative method to measure quasar lifetimes at high redshifts that is sensitive to the quasar's emission in the direction \textit{perpendicular} to our line of sight.
This method relies on the imprints of quasars' radiation on a much smaller ($\sim$kpc) spatial scale compared to the PZs ($\sim$Mpc), namely on the extended nebular emission of the gas inside the circumgalactic medium (CGM) around the quasar. Assuming that this nebular glow is powered by the quasar's radiation, which has been shown to be the case at least for the brightest nebulae \citep{cantalupo_cosmic_2014,borisova_ubiquitous_2016,costa_agn-driven_2022}, the quasar lifetime is measured as the light travel time between the quasar and the outer edge of the nebula. Such a measurement relies on a fundamentally different physical mechanism than PZ-based lifetimes (i.e. light travel timescale vs.\ hydrogen recombination timescale) and thus constitutes an independent measurement of quasar ages. This method has been previously applied to quasars at lower redshifts \citep{trainor_constraints_2013, cantalupo_cosmic_2014, hennawi_quasar_2015, borisova_constraining_2016}, but such extended nebular emission has been found to be ubiquitous around quasars across cosmic time \citep[e.g.][]{heckman_spatially_1991,heckman_spectroscopy_1991,christensen_extended_2006,hennawi_quasars_2013,cantalupo_cosmic_2014,martin_intergalactic_2014,arrigoni_battaia_stacked_2016,borisova_ubiquitous_2016,farina_mapping_2017,farina_requiem_2019,drake_ly_2019}.

In this work, we use deep ($>3.5$ h) observations with the Very Large Telescope/Multi-Unit Spectroscopic Explorer \citep[MUSE,][]{bacon_muse_2010} to measure nebular lifetimes of a sample of $z\sim 6$ quasars with small PZs and thus short inferred lifetimes of $t_{\rm Q} \lesssim 10^6$ yr \citep[some of which are as low as $t_{\rm Q} \sim 10^3$ yr;][]{eilers_detecting_2020,eilers_detecting_2021,yue_detecting_2023}. We chose this sample for two reasons. First, the extremely short PZ-based lifetimes of these high-redshift quasars, taken together with their measured single-epoch BH masses, pose the greatest challenge to our understanding of SMBH assembly. Second, \change{because the growth of ionized nebulae is limited by the speed of light (see Appendix \ref{app:growth}) and the ionizing photon flux decays as $\sim 1/r^{2}$ away from the quasar, these young quasars should be surrounded by nebulae that are small enough to measure their true extent above the noise level.} Notably, \cite{farina_requiem_2019} and \cite{drake_ly_2019} have found an intriguing non-detection of a \lya\ nebula in a deep ($3.7$ h) observation of the $z\sim6$ quasar CFHQS J2100--1715 that exhibits a small PZ corresponding to a lifetime of $t_{\rm Q} \lesssim 10^3$ yr, as well as two more non-detections in shallower observations of quasars with short PZ-based lifetimes, SDSS J0100+2802 and CFHQS J2229+1457. Our study aims to explore this result further and establish whether these quasars have indeed only recently begun their accretion activity, showing only very small (or no) extended nebular emission, or whether very extended nebulae are present, which would point towards sightline-dependent obscuration effects as the cause for the small observed proximity zones. 

We first describe the dataset in \S~\ref{sec:data} and subsequently explain how we search for \lya\ nebulae in \S~\ref{sec:nebulae}. In \S~\ref{sec:lifetimes}, we explain the details of the nebular lifetime measurement. Finally, we discuss the comparison to the previously published, PZ-based lifetimes for our quasar sample as well as the implications of our results in \S~\ref{sec:conclusion}. Throughout this paper, we use the flat $\Lambda$CDM cosmology with $h = 0.67$, $\Omega_M=0.31$, $\Omega_\Lambda=0.69$ \citep{planck_collaboration_planck_2020}.

\begin{table*}[t!]
\centering
\caption{The quasar sample used in this study.}\label{tab:sample}
\begin{tabular}{lcccccccc}
\hline\hline
Quasar & R.A. & Dec. & Redshift & $M_{1450}^a$ & $R_{\rm p}^b$ & $\log t_{\rm Q}^c$ & Ref.$^d$ & Total Exp. Time \\
& [hh:mm:ss.ss] & [dd:mm:ss.s] & & [mag] & [pMpc] & [yr] & & [s]\\
\hline
PSO J004+17 & 00:17:34.47 & +17:05:10.7 & $5.8166$ & $-26.01$ & $1.16 \pm 0.15$ & $3.6^{+0.5}_{-0.4}$ & E21 & 14090 \\
SDSS J0100+2802 & 01:00:13.02 & +28:02:25.8 & $6.3270$ & $-29.14$ & $7.12 \pm 0.13$ & $5.1^{+1.3}_{-0.7}$ & D20 & 12681 \\
VDES J0330--4025 & 03:30:27.92 & $-$40:25:16.2 & $6.249$ & $-26.42$ & $1.69^{+0.62}_{-0.35}$ & $4.1^{+1.8}_{-0.9}$ & E21 & 14090 \\
PSO J158--14 & 10:34:46.51 & $-$14:25:15.9 & $6.0685$ & $-27.41$ & $1.95 \pm 0.14$ & $3.8^{+0.4}_{-0.3}$ & E21 & 30540\\
CFHQS J2100--1715$^\dagger$ & 21:00:54.62 & $-$17:15:22.5 & $6.0806$ & $-25.55$ & $0.37 \pm 0.15$ & $2.3 \pm 0.7$ & E21 & 13338 \\
CFHQS J2229+1457 & 22:29:01.65 & +14:57:09.0 & $6.1517$ & $-24.78$ & $0.47 \pm 0.15$ & $2.9^{+0.8}_{-0.9}$ & E21 & 16908 \\
\hline
\end{tabular}
\begin{flushleft}
\textbf{Notes.} All errors are 1$\sigma$ errors. \\
$^a$ The absolute magnitude at rest frame 1450 \AA. \\
$^b$ The proximity zone size. \\
$^c$ The quasar lifetime based on the proximity zone size. \\
$^d$ The reference from which the redshift, $R_{\rm p}$ and $t_{\rm Q}$ measurements are adopted. Note that all redshifts are from [\ion{C}{2}] observations, except for VDES J0330--4025 whose redshift is \ion{Mg}{2} based while accounting for the \ion{Mg}{2}-[\ion{C}{2}] systematic shift. E21 - \cite{eilers_detecting_2021}; D20 - \cite{davies_constraining_2020}. \\
$^\dagger$ Deep MUSE observations of this quasar have been previously analysed by \cite{drake_ly_2019} and \cite{farina_requiem_2019} We include this data in our study for completeness.
\end{flushleft}
\end{table*}

\section{Data}\label{sec:data}

\subsection{Sample of young $z\sim 6$ quasars}

In order to investigate the implications of short PZ-based lifetimes on SMBH growth, we focus on a sample of quasars that have been previously reported to have small PZs and short inferred lifetimes that are unaffected by gravitational lensing \citep{eilers_detecting_2020,eilers_detecting_2021,yue_detecting_2023}. Detailed information on this sample is shown in \cref{tab:sample}.

For our search of extended \lya\ emission, we use archival observations from the VLT/MUSE integral field spectrograph \citep{bacon_muse_2010} in the Wide Field Mode with natural seeing (WFM-noAO). All objects in our sample have deep ($>3.5$ h) observations, which is important when searching for the faint extended emission. PSO J158--14 and VDES J0330--4025 were observed between November 2020 and March 2021 under program 106.215A (PI: Eilers). PSO J004+17 was observed between November 2021 and September 2022 under program 108.222J (PI: Eilers). For SDSS J0100+2802, we combine observations from program 108.222J (PI: Eilers, observed in October and November 2021) and 0101.A-0656 (PI: Farina, observed in August 2018). Observations of CFHQS J2229+1457 are combined from program 108.222J (PI: Eilers, observed between October 2021 and August 2022) and 0103.A-0562 (PI: Farina, observed in July 2019).\footnote{Note that some of the observations ($<1\ {\rm hr}$) of SDSS J0100+2802 and CFHQS J2229+1457 were previously a part of the REQUIEM survey \citep{farina_requiem_2019} and yielded no nebular detections.} Observations of CFHQS J2100--1715 come from August 2016 (program 297.A-5054, PI: Decarli) and have been analysed before by \cite{drake_ly_2019} and \cite{farina_requiem_2019}. We choose to re-analyse the last quasar here for completeness and for consistency with the rest of our sample. The total exposure times for individual quasars are listed in \cref{tab:sample}. 

\subsection{Data reduction}

The data used in this study were reduced using the MUSE Data Reduction Software \citep{weilbacher_design_2012,weilbacher_muse_2014}, which performs bias subtraction, flat fielding, twilight and illumination corrections, and applies wavelength and flux calibration using standard stars. For CFHQS J2100--1715, we use the same archival observations as well as reduction pipeline as in \cite{farina_requiem_2019} and \cite{drake_ly_2019}. Note that on top of the standard reduction pipeline (v2.6), \cite{farina_requiem_2019} implements a number of custom steps, including improvements to flat-fielding and sky subtraction, and a custom absolute flux calibration and astrometry solution. The rest of the observations of our quasar sample were reduced using v2.8.7 of the MUSE pipeline with standard parameters. As voxel-to-voxel correlations may lead to underestimated noise properties of the thus reduced data cubes \citep{bacon_muse_2015}, we rescaled all variance data cubes to match the measured variance of the background \citep{borisova_ubiquitous_2016,farina_mapping_2017,farina_requiem_2019,arrigonibattaia_qso_2019}. In all cases, the reduced data cubes were further run through the Zurich Atmospheric Purge code \citep[ZAP,][]{soto_zap_2016} to further clean the sky line emission. For the subsequent analysis, we use the ZAP-ped data cubes as they are cleaner from artifacts, which is crucial for our search for faint extended line emission.

\begin{figure*}[t!]
    \centering
    \includegraphics[width=0.9\linewidth,trim={0 0.2cm 0 0},clip]{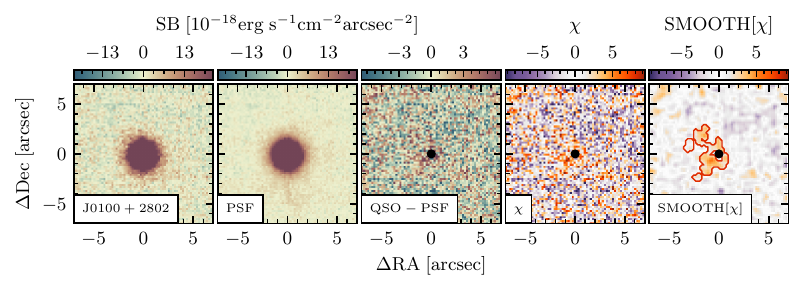}
    \includegraphics[width=0.9\linewidth,trim={0 0.2cm 0 0.2cm},clip]{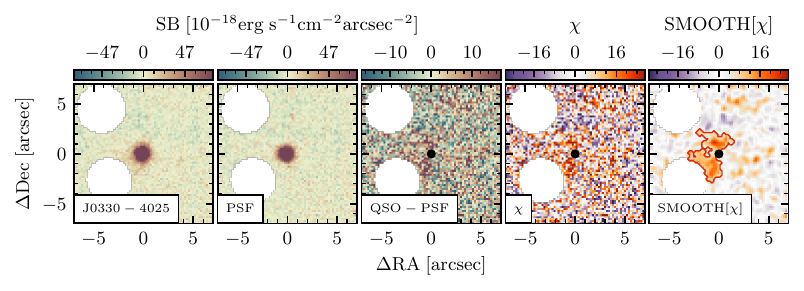}
    \includegraphics[width=0.9\linewidth,trim={0 0.2 0 0.2cm},clip]{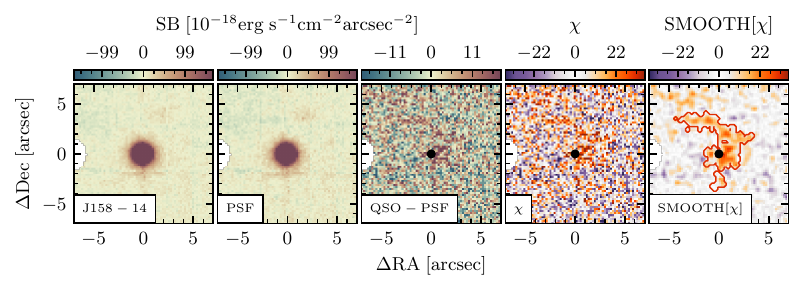}
    \caption{Quasars in our sample with extended nebular emission. From left to right, the panels display pseudo-narrowband images of 1) the quasar's emission around \lya, 2) the PSF extracted from the broad wing of the quasar's \lya\ line, 3) the PSF-subtracted data, 4) the $\chi$ data, essentially representing the SNR of the PSF-subtracted data, and 5) the smoothed $\chi$ data showing the extended nebular emission. Panels 1, 3, 4, and 5 are collapsed across the wavelength range of the detected nebula, and panel 2 is collapsed across the wavelength range of the PSF extraction as displayed in Appendix \ref{app:regions}. White patches correspond to masked foreground sources.}
    \label{fig:qsoswithnebulae}
\end{figure*}

\begin{figure*}[t!]
    \centering
    \includegraphics[width=0.9\linewidth,trim={0 0.2cm 0 0},clip]{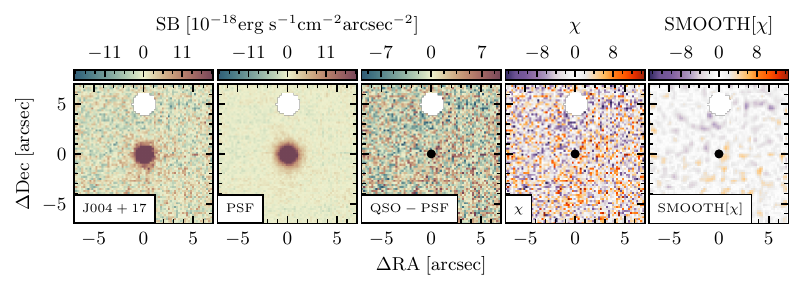}
    \includegraphics[width=0.9\linewidth,trim={0 0.2cm 0 0.2cm},clip]{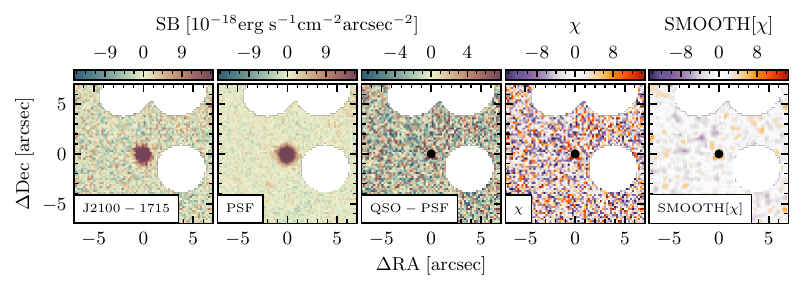}
    \includegraphics[width=0.9\linewidth,trim={0 0.2cm 0 0.2cm},clip]{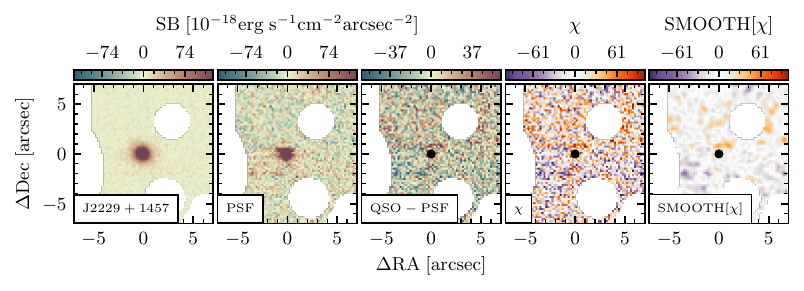}
    \caption{Quasars in our sample without a nebular detection. Panels are the same as in \cref{fig:qsoswithnebulae}, but panels 1, 3, 4, and 5 are now collapsed in a $1000\,{\rm km/s}$ spectral window around the quasar's \lya\ emission.}
    \label{fig:qsoswithoutnebulae}
\end{figure*}

\section{Uncovering \lya\ nebulae}\label{sec:nebulae}

In this section, we describe the analysis pipeline used to search for extended \lya\ emission around the quasars in our sample. This procedure is primarily based on the methods used in the REQUIEM survey \citep{farina_requiem_2019} with a few differences on which we elaborate below.

We first cut the data cubes to $14''\times14''$ around the quasar (corresponding to $80\ {\rm pkpc}\times 80\ {\rm pkpc}$ at $z\sim 6$) to ease computation in subsequent steps. We then mask foreground sources using emission bluewards of the \lya\ wavelength of the quasar where most of the quasar radiation is suppressed due to the IGM \citep{gunn_density_1965}. We perform this step by collapsing the data cube in the foreground of the quasar's \lya\ emission and searching for emission above a signal-to-noise (SNR) threshold (ranging from $2$ to $10$ depending on the specific quasar field) to pick out obvious sources (omitting any residual transmission of the quasar itself). The foreground here is defined to lie between $6000\ {\rm \AA}$ in the observed frame and the wavelength corresponding to $5\times$ the PZ size blueward of \lya\ (as given by \cref{tab:sample}; see Appendix \ref{app:regions}). 

We proceed by masking spectral channels with large surface brightness uncertainty, strong sky line emission or residual detector artifacts. To remove high-uncertainty channels, we use the variance extension of the data cube and mask spectral channels with outlying overall uncertainty. As for the latter two, elevated background is a telltale sign of channels with sky line emission or residual artifacts. We remove these by calculating the background in an annular aperture at a radius comparable to the size of our cutout cube, and we mask spectral channels that show rapid departures from the smoothly varying sky background.

\begin{figure*}[t!]
    \centering
    \includegraphics[width=0.9\linewidth,trim={0 0.3cm 0 0},clip]{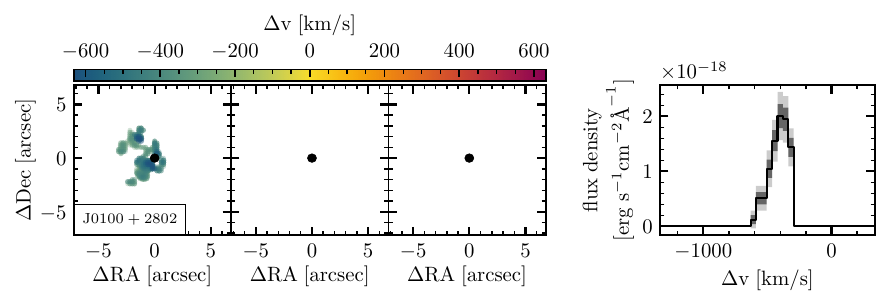}
    \includegraphics[width=0.9\linewidth,trim={0 0.3cm 0 0},clip]{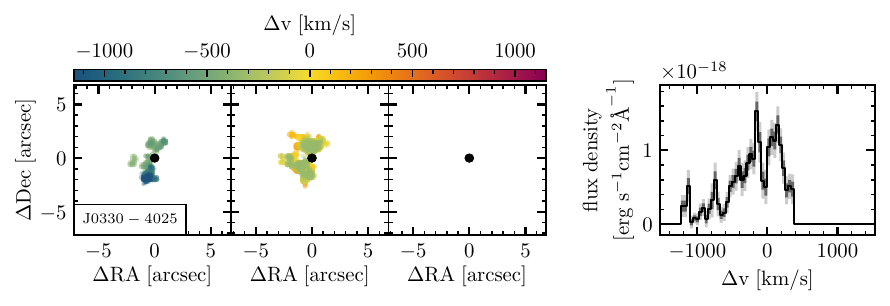}
    \includegraphics[width=0.9\linewidth,trim={0 0.3cm 0 0},clip]{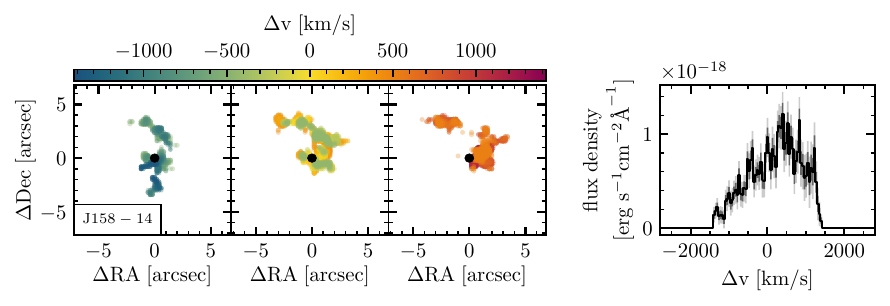}
    \caption{Spectral decomposition of the three nebular detections. The left three panels display the detected nebulae in three spectral channels of equal widths ($766\ {\rm km/s}$ for J0330--4025, $946\ {\rm km/s}$ for J158--14, and $420\ {\rm km/s}$ for J0100+2802), whereby the middle panel is centered on the quasar's systemic redshift as given by \cref{tab:sample}. We also color-code each displayed voxel by the velocity offset from the quasar's \lya\ emission. The right panel shows the integrated emission line profile of the detected nebula.}
    \label{fig:velocity}
\end{figure*}

Subsequently, we perform point-spread-function (PSF) extraction and subtraction using standard methods, summarized here briefly. We extract the PSF from the red wing of the \lya\ line, which originates from the pc-scale broad line region (BLR) of the quasar and thus represents a true unresolved point source. Specifically, we collapse each data cube across a narrow wavelength range at $2500\ {\rm km/s}$ redward of the quasar \lya\ line to create a narrowband image that constitutes our PSF model (displayed in the second panel from the left in \cref{fig:qsoswithnebulae,fig:qsoswithoutnebulae}; see Appendix \ref{app:regions} for the exact spectral regions for each quasar). This wavelength region should thus be free of significant extended emission, assuming the host galaxy contribution is negligible at these redshifts in ground-based observations. With this PSF model at hand, we subtract it from each channel separately, channel-by-channel normalizing the model to the integrated flux inside an aperture with a radius of $0.4''$ centered on the quasar (middle panel in \cref{fig:qsoswithnebulae,fig:qsoswithoutnebulae}).

As the last step, we use the PSF-subtracted cube to search for extended \lya\ emission around the quasar. We first compute a smoothed cube, following \cite{hennawi_quasars_2013,arrigoni_battaia_deep_2015,farina_mapping_2017,farina_requiem_2019},
\begin{equation}
    {\rm SMOOTH}[\chi_{x,y,\lambda}] = \frac{{\rm CONVOL}[{\rm DATA}_{x,y,\lambda}-{\rm PSF}_{x,y,\lambda}]}{\sqrt{{\rm CONVOL}^2[\sigma^2_{x,y,\lambda}]}},
\end{equation}
where ${\rm DATA}_{x,y,\lambda}$ represents the data cube, ${\rm PSF}_{x,y,\lambda}$ is the aforementioned 3-dimensional PSF model normalized at each spectral channel, and $\sigma^2_{x,y,\lambda}$ represents the variance data cube. The ${\rm CONVOL}$ operation denotes a convolution with a 3-dimensional Gaussian kernel with $\sigma_{x,y}=0.2''$ in the spatial direction and $\sigma_\lambda=2.5\ {\rm \AA}$ in the spectral direction \citep[same as in][]{farina_requiem_2019}. The thus constructed ${\rm SMOOTH}[\chi_{x,y,\lambda}]$ is essentially a smoothed SNR cube (pseudo-narrowband images of $\chi_{x,y,\lambda}$ and ${\rm SMOOTH}[\chi_{x,y,\lambda}]$ are shown in the two rightmost panels in \cref{fig:qsoswithnebulae,fig:qsoswithoutnebulae}).

Identifying the extended nebular emission requires us to search the ${\rm SMOOTH}[\chi_{x,y,\lambda}]$ cube for groups of connected voxels that contain significant line emission, and thus to construct a 3D mask of the nebula. We perform this search by first finding the most significant voxel in a $1000 {\rm km/s}$ spectral window around \lya\ and at most $1''$ away from the quasar. Once this voxel is identified, we run a friends-of-friends algorithm to link up all voxels that i) have a significance above a certain ${\rm SNR_{\rm thres}}$, and ii) are within a linking distance $l_{{\rm thres},x,y}$ in the spatial direction and within $l_{{\rm thres},\lambda}$ in the spectral direction. Additionally, we impose that a group has to contain more than $100$ linked voxels to be considered a nebula\footnote{For a cylinder with a base radius of $1''$, $100$ voxels corresponds to a height of $55\ {\rm km/s}$ centered at \lya\ at $z=6$.}.

To explore the robustness of the size of the nebulae, we repeated the linking procedure using the following combination of linking thresholds: ${\rm SNR_{\rm thres}} = \{ 2.0, 3.0 \}$, $l_{{\rm thres},x,y} = \{ 0.4'', 0.6'' \}$, and $l_{{\rm thres},\lambda} = \{ 2.5 {\rm \AA}, 3.75 {\rm \AA} \}$ (corresponding to a linking length of 2 and 3 spatial and spectral pixels, respectively). 
Using the resultant 3D nebular masks, we computed a median mask of the nebula for each quasar, which we use in \cref{fig:qsoswithnebulae,fig:velocity}.

We detect extended \lya\ emission in three quasars in our sample (shown in \cref{fig:qsoswithnebulae}), while the remaining three do not exhibit any nebula above the noise level (\cref{fig:qsoswithoutnebulae}). We further measured the surface brightness limits for all six quasars by collapsing each data cube across five wavelength channels centered at the quasar's \lya\ emission and measuring the variance in a $1{\rm arcsec}^2$ aperture, following the literature \citep[e.g.][]{farina_mapping_2017,farina_requiem_2019}. The measured $5\sigma$ limits are given in \cref{tab:lifetimes} and are comparable across all observations. This fact demonstrates that the non-detections are unlikely to be a consequence of a lack of sensitivity.

Additionally, in \cref{fig:velocity} we visualize the velocity structure in the detected nebulae. We constructed the channel maps shown in the left part of the figure as follows. For each detection, we measured the maximum velocity offset in the nebular mask. We then divided the velocity range defined by this offset into three spectral channels of equal widths ($766\ {\rm km/s}$ for J0330--4025, $946\ {\rm km/s}$ for J158--14, and $420\ {\rm km/s}$ for J0100+2802), such that the second spectral channel is centered on the quasar's \lya\ emission. Within each channel, we color-coded each nebular voxel by its velocity offset from the \lya\ emission of each quasar. Note that despite the large velocity range of all three nebulae, the channel centered on the quasar rest frame encompasses the maximum transverse distance between the quasar and the edge of extended emission, except in the case of J0100+2802 whose nebula is overall blueshifted by $\sim -400\ {\rm km/s}$.

\begin{table*}
    \centering
    \caption{The maximum extent as well as quasar lifetime measurements from the extended \lya\ emission. The uncertainties correspond to the $16$th and $84$th percentiles of the distribution of $d_{\rm max}^{\rm Ly\alpha}$ and $\log{t_{\rm Q}^{\rm Ly\alpha}}$.}\label{tab:lifetimes}
    \begin{tabular}{lccccc}
        \hline\hline
        Quasar & ${\rm SB}^1_{5\sigma,{\rm Ly\alpha}}$ & $\Delta v^{\rm Ly\alpha}$ & $\sigma_v^{\rm Ly\alpha}$ & $d_{\rm max}^{\rm Ly\alpha}$ & $\log{t_{\rm Q}^{\rm Ly\alpha}}$ \\
        & $[{\rm erg\ s^{-1}\ cm^{-2}\ arcsec^{-2}}]$ & $[{\rm km/s}]$ & $[{\rm km/s}]$ & $[{\rm pkpc}]$ & $[{\rm yr}]$ \\
        \hline
        PSO J004+17 & $2.9\times 10^{-18}$ & -- & -- & $<2.37$ & $<3.89$ \\
        SDSS J0100+2802 & $6.5\times 10^{-18}$ & $273_{-63}^{+1692}$ & $70_{-11}^{+471}$ & $17.92_{-3.58}^{+3.58}$ & $4.77_{-0.10}^{+0.08}$ \\
        VDES J0330--4025 & $3.8\times 10^{-18}$ & $793_{-738}^{+775}$ & $184_{-184}^{+187}$ & $8.36_{-6.08}^{+22.83}$ & $4.44_{-0.56}^{+0.57}$ \\
        PSO J158--14 & $3.8\times 10^{-18}$ & $2333_{-765}^{+458}$ & $581_{-164}^{+85}$ & $23.85_{-9.17}^{+23.70}$  & $4.89_{-0.21}^{+0.30}$ \\
        CFHQS J2100--1715 & $2.9\times 10^{-18}$ & -- & -- & $<2.32$ & $<3.88$ \\
        CFHQS J2229+1457 & $4.4\times 10^{-18}$ & -- & -- & $<2.30$ & $<3.88$ \\
        \hline
    \end{tabular}
\end{table*}

Additionally, we extracted the integrated line profile of each median nebula, shown in the right part of \cref{fig:velocity}. We used the nebular line profiles to measure the velocity range, $\Delta v^{\rm Ly\alpha}$, and the velocity dispersion, $\sigma_v^{\rm Ly\alpha}$, for each quasar (\cref{tab:lifetimes}). The properties of the detected nebulae are comparable to nebulae found in large surveys in the literature \citep{borisova_ubiquitous_2016,farina_requiem_2019}. 
 
\section{Quasar lifetime measurements based on the extended Lya nebular emission}\label{sec:lifetimes}

\begin{figure*}[t!]
    \centering
    \includegraphics[width=\linewidth]{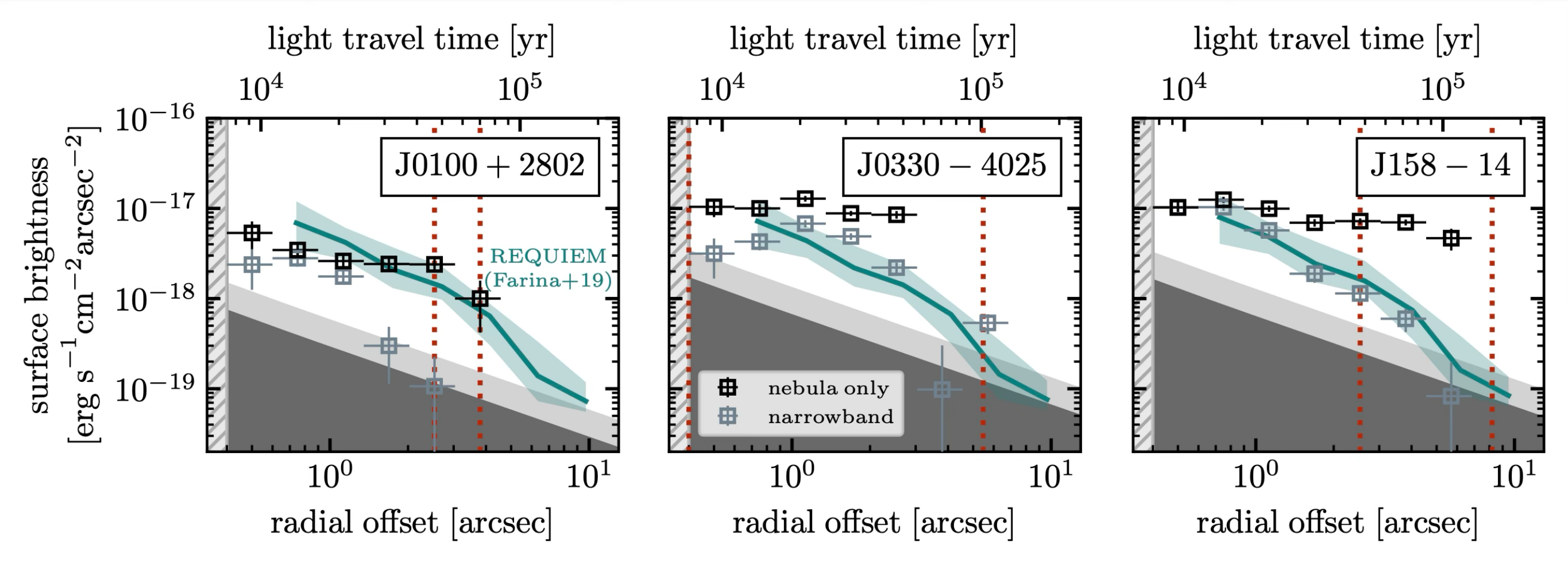}
    \caption{The extracted surface brightness profiles for the three quasars with nebular detections as a function of both the radial offset and the light travel time from the quasar. The black data points correspond to the surface brightness profile of the median nebula, as described in \S~\ref{sec:nebulae}, where we only take into account the nebular pixels so as not to dilute signal at the edge of the nebula. We also show the surface brightness profile calculated via annular averaging of the full narrowband image of the detected nebula as gray data points to illustrate this signal dilution. In the background, we display the median stacked surface brightness profile of $z>5.7$ quasars from \cite{farina_mapping_2017} for comparison. The red dotted lines display the $16$th and $84$th percentile spread of the nebular extent when the nebular search parameters are varied. The quasar lifetime, $t_{\rm Q}^{\rm Ly\alpha}$, is measured as the light travel time from the quasar to the outer edge of the detected nebula in the transverse direction. Note that the horizontal error bars represent the binning of the annular apertures used to calculate the surface brightness. We also display the $1\sigma$ (dark gray) and $2\sigma$ (light gray) background derived from the sky variance in the outermost annular aperture.}
    \label{fig:SB}
\end{figure*}

With the nebular (non-)detections at hand, we proceed to measure the nebular quasar lifetimes, $t_{\rm Q}^{\rm Ly\alpha}$. Assuming that the extended nebular emission is powered primarily by the quasar's radiation \citep[e.g.][]{cantalupo_cosmic_2014,borisova_ubiquitous_2016,costa_agn-driven_2022}, we calculate the quasar lifetime as the light travel time between the quasar and the outer edge of the nebula in the transverse direction. \change{Such calculation only holds in the regime where the size of the nebula grows at or close to the speed of light, which, as we show in Appendix \ref{app:growth}, is valid for lifetimes of at least up to $t_{\rm Q}^{\rm Ly\alpha} \sim 10^5\ {\rm yr}$ for typical CGM densities.}

In order to determine the extent of each nebula, we first compute its surface brightness profile. We collapse the PSF-subtracted data cube across the wavelength range of the detected nebular emission, and use this pseudo-narrowband image of the nebula to perform aperture sum in annular regions centered on the quasar at increasing radii. We choose annular apertures at logarithmically increasing radii, starting at a separation of $0.4''$ away from the quasar -- this is to avoid contamination by the region that was used to normalize the PSF model at each wavelength channel. Because we are interested in measuring the maximum extent of the nebula, we only apply the aperture sum on the voxels contained in the nebular mask to avoid diluting the signal at the outskirts of the nebula. This is in contrast to computing the aperture sum within the whole annulus irrespective of the nebular mask, as is done elsewhere in the literature, e.g. \cite{borisova_ubiquitous_2016,farina_requiem_2019}, which leads the surface brightness profile to smoothly approach the background noise. This difference is illustrated in \cref{fig:SB} for the three nebular detections in our sample, where we show both the nebula-only surface brightness profiles (in black) alongside the full surface brightness profiles (in gray; the difference is also nicely illustrated in e.g. \citeauthor{arrigonibattaia_qso_2019} \citeyear{arrigonibattaia_qso_2019}). Additionally, we show the median surface brightness stack from the REQUIEM survey \citep{farina_requiem_2019}, scaled to the relevant quasar redshift, in the background as a teal curve. We also display the background surface brightness noise in gray, which we derive from the outermost annulus that does not include any nebular pixels. Note that the displayed surface brightness profiles are not corrected for cosmological dimming.

\begin{figure*}[t!]
    \centering
    \includegraphics[width=0.9\linewidth]{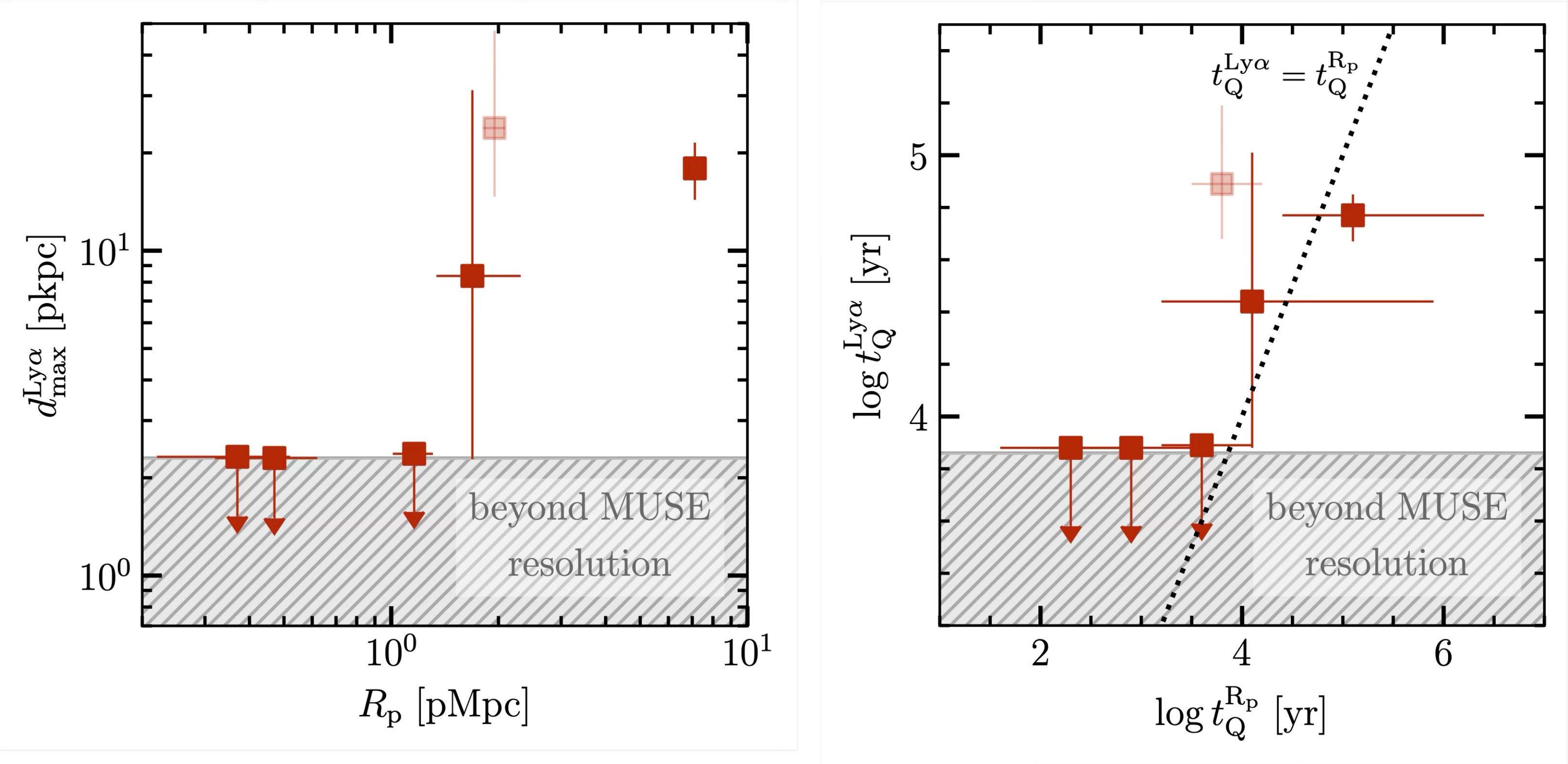}
    \caption{Left: A comparison of the newly measured \lya\ nebula sizes, $d_{\rm max}^{\rm Ly\alpha}$, to the previously published PZ sizes of the quasars in our sample, from \cite{eilers_detecting_2021} and \cite{davies_constraining_2020}. Right: A comparison of the new quasar lifetimes derived from the sizes of \lya\ nebulae, $t_{\rm Q}^{\rm Ly\alpha}$, to the existing PZ-based lifetimes, $t_{\rm Q}^{\rm R_{\rm p}}$ \citep{eilers_detecting_2021,davies_constraining_2020}. Correlations between the nebular and PZ-based quantities suggest that line-of-sight effects are likely not the cause of the short quasar lifetimes observed at high redshift. Indeed, we find agreement in the measured lifetimes for all quasars in our sample, except for one quasar, J158-14 (the faded data point), whose PZ is likely truncated by a proximate absorption system. } 
    \label{fig:tqcomp}
\end{figure*}

We identify the maximum extent of the nebula, $d_{\rm max}^{\rm Ly\alpha}$, based on the largest radius bin with a nebular detection (i.e. based on the black data points in \cref{fig:SB}). We convert this radius from angular units to physical units via the angular diameter distance evaluated at the quasar redshift. 
Subsequently, we convert this maximum nebular extent to a lifetime measurement using the speed of light, $c$,
\begin{equation}
    t_{\rm Q}^{\rm Ly\alpha} = \frac{d_{\rm max}^{\rm Ly\alpha}}{c}.
\end{equation}
Note that $d_{\rm max}^{\rm Ly\alpha}$ here is the distance to the furthest point as projected onto the sky, as opposed to the furthest point in the 3-dimensional space the nebula. The transverse (projected) distance is the relevant quantity for this measurement as position and velocity information along the line of sight are difficult to disentangle -- this is due to peculiar velocities boosting and/or broadening the \lya\ emission profile (i.e. the Finger of God and Doppler effects). Such motion does not bias the nebular extent, and hence the quasar lifetime, measured in the transverse direction.

As detailed in \S~\ref{sec:nebulae}, we performed the nebula search using a range of thresholds for the SNR and the spatial and spectral linking lengths of individual voxels. We use all the thus-created nebular masks to measure a distribution of nebular extents and quasar lifetimes. We include their median measurements as well as the $16$th and $84$th percentile uncertainties in \cref{tab:lifetimes} and also show these as vertical dotted red lines in \cref{fig:SB}. For the quasars in our sample without detected \lya\ nebulae, we place an upper bound on their properties based on the spatial region we do not resolve in our analysis ($0.4''$). Note that the surface brightness limits given in \cref{tab:lifetimes}, which are comparable across all observations included in our sample, suggest that these are true non-detections -- in fact, even if the nebulae of these three quasars were $10\times$ fainter than our detections \citep{mackenzie_revealing_2021}, we would still expect to marginally detect them at this sensitivity.

\section{Implications for SMBH growth}\label{sec:conclusion}

We have presented new lifetime measurements for a sample of six quasars at $z\sim6$ based on the spatial extent of their nebular \lya\ emission in deep ($>3.5 {\rm h}$) MUSE observations with comparable sensitivity. Three of these quasars exhibit extended \lya\ emission, while the nebulae of the remaining three quasars remain undetected.

This quasar sample is particularly intriguing to study from the perspective of lifetimes, as these quasars all exhibit extremely small PZs ($R_{\rm p}$) and thus very short PZ-based quasar lifetimes \citep[$t_{\rm Q}^{\rm R_{\rm p}}$, summarized in \cref{tab:sample};][]{eilers_detecting_2020,eilers_detecting_2021,davies_constraining_2020}. A comparison between the sizes of the \lya\ nebulae of these quasars, $d_{\rm max}^{\rm Ly\alpha}$, \change{including the non-detections}, and their PZs (shown in the left panel of \cref{fig:tqcomp}) \change{suggests a correlation, meaning that line-of-sight} effects and time-variable obscuration are unlikely the cause of these extremely small PZs. Indeed, our new nebular lifetime measurements (shown in the right panel of \cref{fig:tqcomp}) \change{are also consistent} with these short quasar lifetimes based on PZs. This is remarkable particularly because the two lifetime measurement methods are based on very different spatial scales (the kpc-scale CGM vs the Mpc-scale IGM) and very different physical mechanisms (light travel timescales vs ionization and recombination timescales), and thus constitute two truly independent measurements.

One of the quasars in our sample, J158--14, is shown as a faded data point in \cref{fig:tqcomp}. This is because, upon a closer inspection of the MUSE data, we identified a proximate \lya\ emitter (LAE) that seems to be truncating the line-of-sight PZ of this quasar, and thus results in an underestimate of the $t_{\rm Q}^{\rm R_{\rm p}}$ measurement. This reconciles the tension between the two lifetime measurements for this quasar, whereby its nebular lifetime shows that this quasar is substantially older than its PZ seems to suggest. Coincidentally, this foreground LAE imparts no metal absorption lines on the spectrum of this quasar and is presented in a companion paper \citep{durovcikova_extremely_2025}.

The main caveats of these nebular lifetime measurement are as follows. First, we operate under the assumption that we are indeed seeing the edge of the nebula in the surface brightness profiles in \cref{fig:SB}. Since we chose to compute our surface brightness profiles so as to avoid diluting the signal at the outskirts of the detected nebula, we would expect the surface brightness profiles to smoothly approach the background noise level if these observations were not sensitive enough to capture the full extent of the nebular emission. This is indeed supported by noticing that, in \cref{fig:SB}, the full narrowband surface brightness profiles of J0100+2802 and J0330-4025\footnote{Note that the last gray data point for J0330-4025 in \cref{fig:SB} is slightly elevated; this is due to a few higher-SNR patches in the field that have not been identified to be a part of this quasar's nebula, see middle panel of \cref{fig:qsoswithnebulae}.} (shown as gray data points), show a steeper downturn towards the noise level than the median stack of $z> 5.7$ quasars from \cite{farina_requiem_2019} shown in teal. In the case of J158--14, the surface brightness profile seems to suggest that the extent of the nebula might continue below the noise level, which is not surprising given that this quasar seems to have a much longer lifetime than the lifetime revealed by its PZ.

Additionally, this method assumes that the dominant mechanism behind the observed extended \lya\ emission is the recombination of a free electron with an ionized hydrogen atom. This assumption has been supported by studies at lower-redshift \citep{leibler_detection_2018,langen_characterizing_2023} that compared the extended \lya\ emission to the extended \ha\ emission around quasars and found a close agreement \change{with recombination based on} their respective line fluxes. However, it should be noted that \lya\ can in general also arise from resonant scattering which can alter the velocity profile of its emission. The presence of scattering as the \lya\ photons make their way from the BLR to and through the CGM would thus result in an increased light travel time that we are currently not accounting for. \cite{costa_agn-driven_2022} has shown that resonant scattering increases the asymmetries and also the velocity dispersion in the nebular \lya\ line profiles. Our measured velocity dispersions (\cref{tab:lifetimes}) are in line with the values expected for recombination radiation from \cite{costa_agn-driven_2022}, although the asymmetries and low spectral resolution of the line profiles shown in \cref{fig:velocity} make it difficult to draw any definite conclusions. This limitation motivates further deep integral-field-unit observations of the rest-frame optical lines where resonant scattering is suppressed (e.g. \ha, \oiii) to enable a more complete picture of nebular quasar lifetimes.

In summary, our results show that the nebular quasar lifetimes from extended \lya\ emission \change{are consistent with the existing PZ-based lifetimes, assuming  recombination radiation is the dominant emission mechanism}. This result disfavors the scenario where time-variable obscuration effects would have led to short measured UV-luminous quasar lifetimes.
This finding provides further support that these $z\sim 6$ quasars could have started their accretion only recently, but a study of their rest-frame optical emission is needed for more conclusive evidence and investigation into alternative SMBH growth pathways.

\section*{Acknowledgments}

\change{We would like to thank the referee for the comments that improved the quality of this manuscript.} We would also like to thank Roberto Decarli and Bram Venemans for helpful discussions.

E.P.F. is supported by the international Gemini Observatory, a program of NSF NOIRLab, which is managed by the Association of Universities for Research in Astronomy (AURA) under a cooperative agreement with the U.S. National Science Foundation, on behalf of the Gemini partnership of Argentina, Brazil, Canada, Chile, the Republic of Korea, and the United States of America. R.A.M acknowledges support from the Swiss National Science Foundation (SNSF) through project grant 200020\_207349. \change{C.M. acknowledges support from Fondecyt Iniciacion grant 11240336 and the ANID BASAL project FB210003}.

Based on observations collected at the European Southern Observatory under ESO programs: 106.215A, 108.222J, 0101.A-0656, 0103.A-0562, and 297.A-5054.

%

\vspace{5mm}




\newpage
\appendix

\section{How fast do \lya\ nebulae grow?}\label{app:growth}

\change{The method presented in \S~\ref{sec:lifetimes} uses the light crossing time of the ionized nebula as the relevant timescale to convert the size of the nebula to a quasar lifetime constraint. This works if the \lya\ nebula itself grows at or close to the speed of light when the quasar radiation turns on. In what follows, we demonstrate that this is the case for typical CGM conditions around quasars and for timescales relevant to this work.}

\change{The CGM is known to be a highly inhomogeneous, multi-phase medium. \lya\ nebulae are primarily thought to be powered through the hydrogen recombination cascade \citep{leibler_detection_2018,langen_characterizing_2023} and trace the denser, colder neutral gas clouds within the CGM, with typical number densities of $n_{\rm H} \approx 1\ {\rm cm}^{-3}$ \citep{cantalupo_cosmic_2014, arrigoni_battaia_deep_2015,hennawi_quasar_2015}. However, the volume filling factor of these neutral gas clouds is generally small, $C_{\rm V} < 10^{-2}$ \citep{mccourt_characteristic_2018,pezzulli_high_2019}, and can be as low as $C_{\rm V} \sim 10^{-5}-10^{-4}$ \citep{prochaska_quasars_2009}. Such tiny values of $C_{\rm V}$ imply that most of the gas in the CGM is ionized, enabling the ionizing photons to travel at the speed of light until they encounter a neutral patch along their line of sight. In other words, the observed \lya\ nebulae simply trace the sightlines that illuminate these small, dense neutral gas clouds, and the light travel time to the farthest illuminated gas cloud should thus accurately reflect the quasar lifetime.}

\begin{figure}
    \centering
    \includegraphics[width=\linewidth]{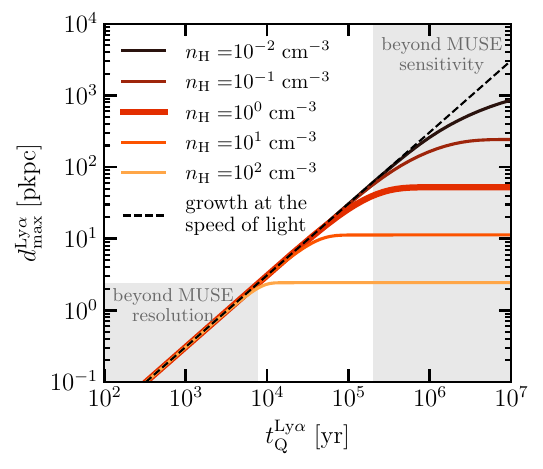}
    \caption{\change{The growth of the ionized \lya\ nebula size, $d_{\rm max}^{\rm Ly\alpha}$, as a function of quasar lifetime, $t_{\rm Q}^{\rm Ly\alpha}$, following the relativistic treatment of \cite{shapiro_relativistic_2006} for a range of hydrogen number densities, $n_{\rm H}$. For typical densities of $n_{\rm H} \approx 1\ {\rm cm}^{-3}$ (thick curve) that are responsible for \lya\ nebulae \citep{cantalupo_cosmic_2014, arrigoni_battaia_deep_2015,hennawi_quasar_2015}, the expansion of the nebula size closely follows the speed of light (dashed line) up to $t_{\rm Q}^{\rm Ly\alpha} \sim 10^5\ {\rm yr}$. This is actually a lower bound on the range of lifetimes our method is valid for, as in a more realistic, mostly ionized CGM with a low volume filling factor of neutral gas clouds, the ionization front would move close to the speed of light out to even larger distances.}}
    \label{fig:nebgrowth}
\end{figure}

\change{One might still object that, even though the photons travel at the speed of light through the CGM until they hit a neutral gas cloud, the time it takes to photoionize this residual neutral gas could be limiting the speed at which nebulae can grow. To address this concern, we proceed to demonstrate that the ionization front travels at or close to the speed of light through gas under conditions relevant for \lya\ emission and for timescales relevant to this work.}

\change{We consider a toy model where we assume that the CGM around the quasar is homogeneous and only composed of neutral hydrogen gas at a constant number density $n_{\rm H}$. Note that this simple model will yield the most pessimistic estimate on how fast the ionization front progresses through the CGM due to the increased rate of recombinations in a fully neutral medium. Following the treatment of relativistic ionized fronts in \cite{shapiro_relativistic_2006} (Eqs. 30--36 therein), we compute the size of the ionized nebula, $d_{\rm max}^{\rm Ly\alpha}$, around a typical quasar of luminosity $L=10^{47}\ {\rm erg/s}$ as a function of the quasar lifetime, $t_{\rm Q}^{\rm Ly\alpha}$, for a range of hydrogen number densities $n_{\rm H}$ typically considered in relevant literature \citep[e.g.][]{arrigoni_battaia_deep_2015}. We plot the results in \cref{fig:nebgrowth} as solid curves at a range of colors (the thick curve highlighting the typical value $n_{\rm H} = 1\ {\rm cm}^{-3}$) alongside a dashed line depicting growth at the speed of light.}

\change{From \cref{fig:nebgrowth}, we can see that even if the quasar had to carve out an ionized bubble in this unrealistic fully neutral CGM model, the growth of this ionized region would still progress close to the speed of light up until $t_{\rm Q}^{\rm Ly\alpha} \sim 10^5\ {\rm yr}$ from the moment the quasar has turned on. This result implies that for all of our measurements in this manuscript, the light travel time to the edge of the nebula is indeed a good tracer of the actual quasar lifetime, especially given that the quasar only needs to photoionize small neutral gas clouds instead of the whole CGM.}

\change{Two further points can be made here. First, note that the lifetime of $t_{\rm Q}^{\rm Ly\alpha} \sim 10^5\ {\rm yr}$ also corresponds roughly to the limit of what we can constrain with our observations, as the predicted surface brightness profile would fall below the sensitivity achieved with MUSE for longer lifetimes (see \cref{fig:SB}). Second, in the parameter space that is inaccessible due to the limited spatial resolution of MUSE, the nebula would grow close to the speed of light for all $n_{\rm H}$ values considered in our toy model. This means that we can actually place an upper bound on the lifetimes of quasars whose \lya\ nebulae are not detected in deep MUSE observations, as, even if their growth were limited by the speed of the ionization front, their nebulae should be expanding at the speed of light even in the densest parts of the CGM.}

\section{Spectral regions}\label{app:regions}

\begin{figure*}[h!]
    \centering
    \includegraphics[width=0.9\linewidth,trim={0 0.65cm 0 0.2cm},clip]{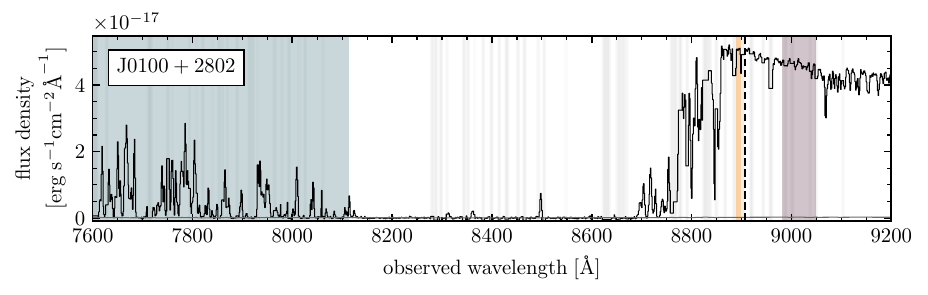}
    \includegraphics[width=0.9\linewidth,trim={0 0.65cm 0 0.2cm},clip]{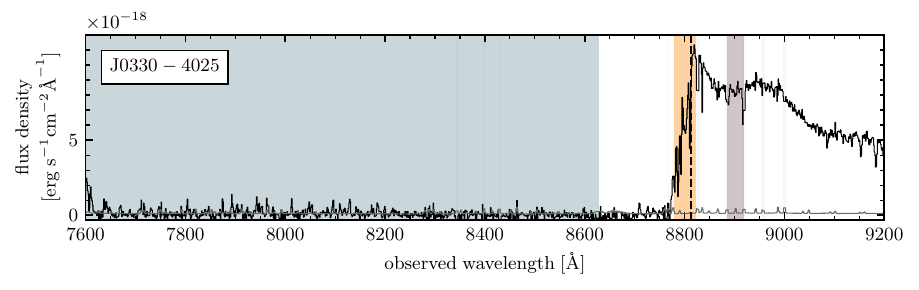}
    \includegraphics[width=0.9\linewidth,trim={0 0.2cm 0 0.2cm},clip]{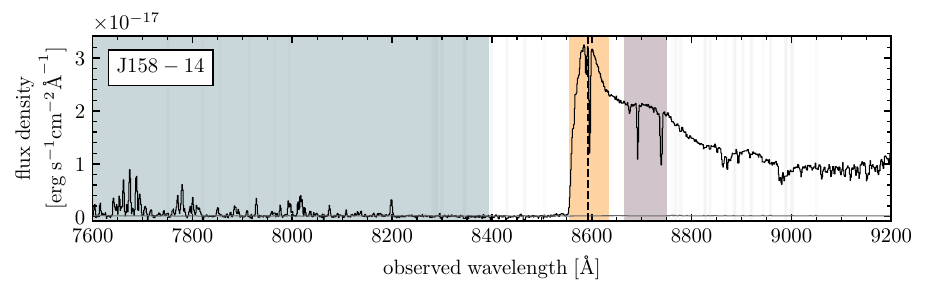}
    \caption{Spectral regions corresponding to the wavelengths where foreground sources are identified (in blue; foreground defined as $5\times$ the size of the quasar's PZ blueward of \lya), where PSF is extracted (in purple), and where nebular emission is detected (in orange). The black data in each row shows the quasar spectrum extracted from the MUSE data cube using an aperture with a radius of three pixels, with the error vector displayed in gray. The vertical dashed line marks \lya\ corresponding to the quasar's systemic redshift given in \cref{tab:sample}, and the faint gray vertical lines show the spectral channels that have been masked as described in \S~\ref{sec:nebulae}.}
    \label{fig:spectralwithnebulae}
\end{figure*}

\begin{figure*}[h!]
    \centering
    \includegraphics[width=0.9\linewidth,trim={0 0.65cm 0 0.2cm},clip]{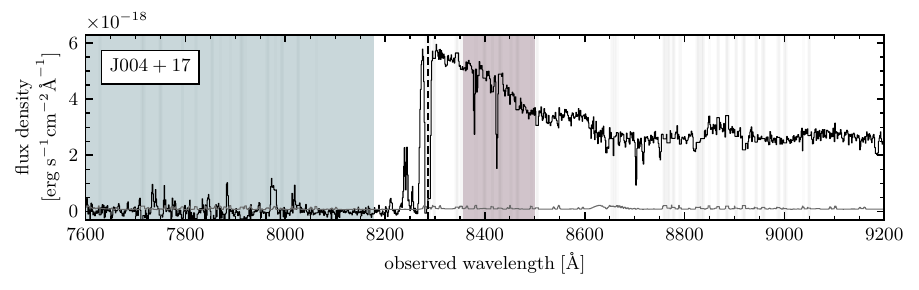}
    \includegraphics[width=0.9\linewidth,trim={0 0.65cm 0 0.2cm},clip]{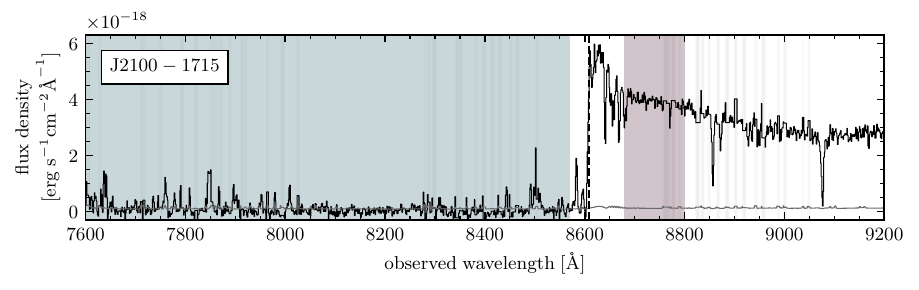}
    \includegraphics[width=0.9\linewidth,trim={0 0.2cm 0 0.2cm},clip]{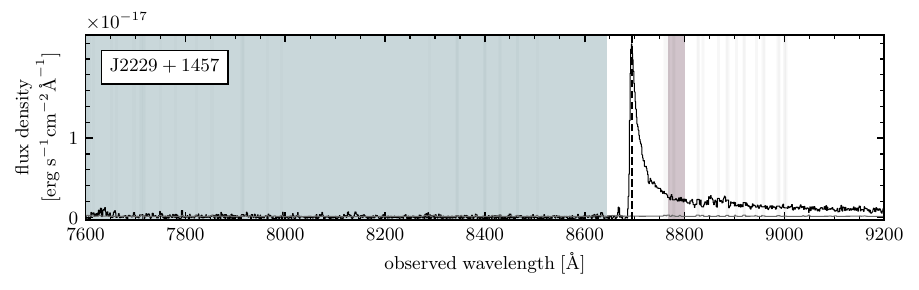}
    \caption{Same as \cref{fig:spectralwithnebulae}, but for quasars without a nebular detection.}
    \label{fig:spectralwithoutnebulae}
\end{figure*}

In \cref{fig:spectralwithnebulae,fig:spectralwithoutnebulae}, we display the extracted spectrum for each quasar in our sample as well as the spectral regions that have been used to search for foreground sources (in blue), to extract the PSF (in purple), and, if detected, the region where extended nebular emission was found (in orange; corresponding to the median nebula as described in \S~\ref{sec:nebulae}). The vertical dashed line marks the \lya\ emission of the quasar corresponding to the systemic redshift from \cref{tab:sample}. These figures also show the spectral channels that have been masked due to having large surface brightness uncertainty, strong sky line emission or residual detector artifacts (as described in \S~\ref{sec:nebulae}; displayed as faint vertical gray lines).

\bibliography{refs}{}
\bibliographystyle{aasjournal}



\end{document}